\documentclass[aps,prc,floatfix,groupedaddress,showpacs,amsfonts, twocolumn]{revtex4}
\usepackage{bm}
\usepackage{epsfig}
\usepackage{amsmath}
\usepackage{dcolumn}

\begin{document}
\title{Elastic scattering of $^{17}$F, $^{17}$O, and $^{19}$F on a heavy target in a microscopic continuum discretized coupled-channels method}

\author{J. Grineviciute}
\affiliation{Physique Nucl{\'e}aire Th{\'e}orique et Physique Math{\'e}matique, C.P. 229, Universit{\'e} Libre de Bruxelles (ULB), B 1050 Brussels, Belgium}

\author{P. Descouvemont}
\affiliation{Physique Nucl{\'e}aire Th{\'e}orique et Physique Math{\'e}matique, C.P. 229, Universit{\'e} Libre de Bruxelles (ULB), B 1050 Brussels, Belgium}

\begin{abstract}
\begin{description}
\item[Background:] In the traditional continuum discretized coupled-channels (CDCC) method, the clusters of the projectile are structureless. Exotic nuclei exhibit unusual properties and often show significant couplings to the continuum. Therefore, reaction models that consider a more accurate structure of the projectile are often preferable. Microscopic description of the projectile, based on an effective nucleon-nucleon (NN) interaction, in a microscopic CDCC (MCDCC) model [Descouvemont and  Hussein, Phys. Rev. Lett. {\bf 111}, 082701 (2013)] has been successfully applied to the $^{7}$Li+$^{208}$Pb scattering.

\item[Purpose:] The MCDCC method is applied to low-energy elastic scattering of $^{17}$F, $^{17}$O, and $^{19}$F on $^{58}$Ni and $^{208}$Pb targets. The goal of the calculations is twofold: to test the adequacy and the accuracy of the MCDCC model for heavier projectiles and to study the contribution of various channels to the elastic scattering cross sections.

\item[Methods:] The elastic scattering cross sections are calculated by using the MCDCC method. The nucleon-target optical potential is folded with the projectile densities resulting from an effective NN interaction, which includes central nuclear, spin-orbit and Coulomb terms. Discretization of the continuum is achieved via the pseudostate method. Coupled equations are solved by using the $R$-matrix method on a Lagrange mesh.

\item[Results:] For the test case of $^{17}$F at 10 MeV/nucleon, the cross sections are weakly sensitive to the choice of the effective NN interaction. Three different energy-dependent optical nucleon-target potentials provide a similar reasonable agreement with data. Just below the Coulomb barrier, the MCDCC significantly underestimates the cross sections at larger angles. The coupling to continuum is not significant in most of the assessed cases.

\item[Conclusions:] The MCDCC is very satisfactory in the sense, that it includes the microscopic properties of the projectile in a reaction model. Well above the Coulomb barrier, the cross sections are in a good agreement with the data. The reasons for the discrepancy between the data and the calculated cross sections at the lower energies, which is also observed in a traditional CDCC, are unclear.

\end{description}
\end{abstract}


\maketitle

\section{Introduction}
The continuum discretized coupled-channels (CDCC) method is a well-established theory \cite{Ya12,Dr10,KuMo12} for direct nuclear reactions, including breakup. In the traditional nonmicroscopic CDCC, the nucleus can be broken into two fragments in the nuclear or Coulomb field of the target, thus involving three particles in the final state. The clusters are structureless and interact via phenomenological potentials. The three-body scattering problem is solved through a set of coupled Schr{\"o}dinger-like differential equations. The breakup is included by expanding the full three-body wave function in terms of a complete basis of the projectile's bound and continuum states. The coupling to continuum states is carried out through a discretization by using pseudo states, averaging over energy, or  by using a midpoint method \cite{Ya12}.

The CDCC method has been extended to a fully microscopic treatment of the two-cluster projectile and successfully applied to the $^{7}$Li+$^{208}$Pb scattering \cite{De13}. Rather than traditionally using the target-first cluster and the target-second cluster optical potentials, the model relies on a nucleon-target optical potential, which is folded with the projectile densities, resulting from a Hamiltonian based on an effective nucleon-nucleon (NN) interaction. A microscopic description of the projectile is especially desirable for reactions involving exotic nuclei, as they present halo and cluster structure phenomena. Halo states exist as ground states of loosely bound nuclei as well as excited states of well-bound nuclei. Due to the proximity to continuum and to the large root mean square (rms) radii, these nuclei exhibit significant couplings to the continuum. Such nuclei are studied experimentally through reactions induced by radioactive beams. Due to the fragility of the exotic nuclei, as well as low production cross sections, these studies are often difficult. Provided a reliable nucleon-target interaction is available, the microscopic CDCC (MCDCC) model is a powerful predictive tool, when experimental data are not available.

The current work presents an application of MCDCC to the low-energy elastic scattering of A=17 mirror nuclei, as well as $^{19}$F on a heavy target. $^{17}$F is a proton drip line nucleus with its $5/2^{+}$ valence proton bound by 0.6 MeV. The first excited state of $^{17}$F is a proton-halo state \cite{Mo97}. Its mirror nucleus $^{17}$O is well bound (S$_{n}=4.1$ MeV). $^{19}$F  is a stable nucleus with $S_{\alpha}=4.014$ MeV and $S_{p}=7.994$ MeV. Microscopic studies have been previously performed for the structure of $^{17}$F \cite{BaDe98} and $^{19}$F \cite{DeBa87}, the elastic scattering of $^{17}$F from a heavy target has been analysed in the traditional CDCC approach \cite{KuMo12,Ma10}.

The goal of the present calculations is to investigate the accuracy of the MCDCC model for heavier projectiles and to study the contribution of various channels to the elastic scattering cross sections. The MCDCC is tested for the $^{17}$F, $^{17}$O, and $^{19}$F projectiles on the $^{58}$Ni and $^{208}$Pb targets at energies around the Coulomb barrier. The $^{17}$F and $^{17}$O  projectiles are described as a composition of $^{16}$O + nucleon subsystems and $^{19}$F nucleus as a composition of $^{15}$N + $\alpha$.
\section{Formalism}
\subsection{Projectile description}
\subsubsection{Schr{\"o}dinger equation}
The Hamiltonian
\begin{align}
& H_{0}\left(\textbf{r}_{1},\ldots,\textbf{r}_{A}\right) =\sum^{A}_{i=1} T_{i} + \sum^{A}_{i>j=1} V_{ij}\left(\textbf{r}_{i}-\textbf{r}_{j} \right)= \nonumber\\
& H_{1}+H_{2}+T_{\rho}+\sum^{A_{1}}_{i=1}\sum^{A_{2}}_{j=1}V_{ij}
\label{H0}
\end{align}
describes the internal structure of the projectile consisting of $A$ nucleons. Here, $T_{i}$ is the kinetic energy operator associated with coordinate $r_{i}$, defined from the c.m. of the projectile, $V_{ij}$ is an effective NN interaction, containing Coulomb, central nuclear and spin-orbit parts. In the MCDCC model, $H_{0}$ is  expressed as a composition of two clusters, although, in principle, more clusters can be included. $H_{1}$ and $H_{2}$ are the internal Hamiltonians of the clusters, $T_{\rho}$ is the relative kinetic energy operator associated with coordinate $\rho$.
The Schr{\"o}dinger equation for the projectile
\begin{equation}
\left(H_{0}-E_{1}-E_{2}\right) \phi^{jm}_{k} =  E^{j}_{k}  \phi^{jm}_{k} 
\label{H0eq}
\end{equation}
is solved as a generalized eigenvalue problem. $E_{1}=\left\langle \phi_{1}\left| H_{1} \right|\phi_{1}\right\rangle$ and $E_{2}=\left\langle \phi_{2}\left| H_{2} \right|\phi_{2}\right\rangle$ are the internal energies of the two clusters, and $k$ is the excitation level. The projectile wave functions $\phi^{jm}_{k}$, based on the internal antisymmetrized cluster wave functions $\phi_{1}$ and $\phi_{2}$ in the resonating group method (RGM), are associated with bound states at $E^{j}_{k} < 0$ and represent square-integrable approximations for continuum wave functions for $E^{j}_{k} > 0$. A discretization of the continuum is performed via the pseudostate method. 

For a $\left(I_{1}I_{2}sljmk\right)$ channel, the RGM wave function is written as
\begin{align}
& \phi^{lsjm}_{k} \left( \bm \xi_{1},\bm  \xi_{2},\bm  \rho \right) = \nonumber\\
& \mathcal{ A } \left[ \left[ \phi^{I_{1}}_{1} \left(\bm \xi_{1}\right) \otimes \phi^{I_{2}}_{2} \left(\bm \xi_{2}\right) \right]^{s} \otimes Y_{l}\left(\Omega_{\rho}\right)i^{l} \right]^{jm} g^{lsj}_{k}\left(\rho\right) , \nonumber\\
& \ \ \Omega\equiv\left(\theta,\varphi\right)
\label{Eq1}
\end{align} 
and $\phi^{jm}_{k}=\sum_{ls} \phi^{lsjm}_{k}$. In Eq. (\ref{Eq1}) $\mathcal{ A }$ is the A nucleon antisymmetrizer. $I_{1}$ and $I_{2}$ are the spins of the two clusters, $\bm \xi_{1}=\left\{\textbf{r}_{i1}-\textbf{R}_{c.m.1}  \right\}$ and $\bm \xi_{2}=\left\{\textbf{r}_{i2}-\textbf{R}_{c.m.2}  \right\}$ are the sets of their internal translationally invariant coordinates. In principle, any number of core excitations can be included. The present calculations use only the ground state wave function for the $^{16}$O cluster in the $^{17}$F and $^{17}$O projectiles and the ground state and the first excited state wave functions for the $^{15}$N cluster in the $^{19}$F projectile. The internal wave functions $\phi_{1}$ and $\phi_{2}$ are defined in the harmonic oscillator model with a common harmonic oscillator size parameter $b$, as to avoid the spurious c.m. components. 

 In the equivalent generator coordinate method (GCM), the relative RGM component $g^{lsj}_{k}$ is expanded over a Gaussian basis
\begin{equation}
 g^{lsj}_{k}\left( \rho\right) =\sum_{n} f^{lsj}_{k}\left(S_{n}\right) \Gamma^{l}_{k}\left( \rho,S_{n}\right) , 
\end{equation}
where \cite{DeDu11}
{\small\begin{equation}
\Gamma^{l}_{k} \left(\rho ,S_{n}\right) = \left(\frac{\mu_{0}}{\pi b^{2}}\right)^{3/4} \exp\left[ -\frac{\mu_{0}}{2b^{2}}\left( \rho^{2}+S^{2}_{n}\right) \right]i_{l}\left( \frac{\mu_{0}\rho S_{n}}{b^{2}}\right) 
\end{equation}}
and $i_{l}\left(x\right)=\sqrt{\pi/2x}I_{l+1/2}\left(x\right)$ is a modified spherical Bessel function, $\mu_{0}=\frac{A_{1}A_{2}}{A}$, allowing the projectile wave function $\phi^{jm}_{k}$ to be expressed as a superposition of the projected Slater determinants $\Phi^{lsjm}$  in a two-center harmonic oscillator shell model with the same harmonic oscillator parameter $b$ \cite{DeDu11}:
\begin{equation}
\phi^{jm}_{k} =\frac{1}{\Phi_{c.m.}}\sum_{lsn} f^{lsj}_{k}\left(S_{n}\right) \Phi^{lsjm}\left(S_{n}\right)  .
\label{Eq2}
\end{equation}
Here, $\Phi_{c.m.}$ is the c.m. wave function \cite{DeDu11}. The expansion coefficients $f^{lsj}_{k}\left(S_{n}\right)$ in Eq. (\ref{Eq2}) are obtained by solving a  generalized eigenvalue problem involving projected GCM kernels:
\begin{align}
& \sum_{n^{\prime}lsl^{\prime}s^{\prime}} \left[ \left\langle\bar{\Phi}^{lsj}\left(S_{n}\right)\left|  H_{0}   \right| \bar{\Phi}^{l^{\prime}s^{\prime}j}\left(S_{n^{\prime}}\right) \right\rangle \right. \nonumber\\
& \left. - E^{j}_{k} \left\langle\bar{\Phi}^{lsj}\left(S_{n}\right) |  \bar{\Phi}^{l^{\prime}s^{\prime}j}\left(S_{n^{\prime}}\right) \right\rangle \right] f^{l^{\prime}s^{\prime}j}_{k}\left(S_{n^{\prime}}\right) =0 .
\end{align}
The matrix elements are calculated between the Slater determinants $\Phi^{lsj}$ and then corrected for the c.m. components. 

\subsubsection{Projectile densities}
The projectile transition densities are used in the MCDCC model to determine the coupled-channel potentials (see Eq. (\ref{Eq12}) below), and are obtained following Ref. \cite{BaDe94}. The transition mass densities, defined as \cite{Ka81}
\begin{equation}
\rho_{j^{\prime}m^{\prime}k^{\prime},jmk}\left(\textbf{r}\right)=\left\langle \phi^{j^{\prime}m^{\prime}}_{k^{\prime}}\left| \sum^{A}_{i=1} \delta\left( \textbf{r}_{i}-\textbf{r} \right) \right|\phi^{jm}_{k} \right\rangle  ,
\label{delta}
\end{equation}
may be expanded in multipoles 
\begin{equation}
\rho_{j^{\prime}m^{\prime}k^{\prime},jmk}\left(\textbf{r}\right)=\sum_{\lambda} \rho^{\left( \lambda\right)}_{j^{\prime}j}\left(r\right)  \left\langle j m \lambda \mu |j^{\prime} m^{\prime}  \right\rangle  Y^{*}_{\lambda\mu} \left( \Omega_{r} \right) .
\label{Eqmulti}
\end{equation}
The corresponding proton and neutron density matrix elements are defined as \cite{BaDe94}
{\small\begin{equation}
\rho^{p (n)}_{j^{\prime}m^{\prime}k^{\prime},jmk}\left(\textbf{r}\right)=\left\langle \phi^{j^{\prime}m^{\prime}}_{k^{\prime}}\left| \sum^{A}_{i=1} \delta\left( \textbf{r}_{i}-\textbf{r} \right) \left(\frac{1}{2}-\epsilon^{p (n)} t_{zi} \right)  \right|\phi^{jm}_{k} \right\rangle 
\label{Eqdens}
\end{equation}}
and can be expanded analogously to Eq. (\ref{Eqmulti}). Here, the convention of $t_{z}=1/2$  for the neutron is used, and $\epsilon^{p (n)}$ equals +1 and -1 for proton and neutron densities, respectively. The density matrix element (\ref{Eqdens}) between the two Slater determinants $\Phi^{lsjm}$ contains a spurious c.m. component, which can be factored out from the form factors using the expressions given in Ref. \cite{BaDe94}. 

\subsection{Projectile-target system}

The Hamiltonian for an inert target and a projectile  
\begin{equation}
H = H_{0}\left(\textbf{r}_{1},\ldots,\textbf{r}_{A}\right) + T_{R} + \sum^{A}_{i=1} V_{ti}\left(\textbf{r}_{i}-\textbf{R} \right) 
\label{totalH}
\end{equation}
involves the projectile Hamiltonian $H_{0}$ (\ref{H0}), a kinetic energy operator for the target-projectile relative motion $T_{R}$ and a target-nucleon interaction $V_{ti}$, where $R$ is the target-projectile coordinate. The total wave function of the projectile-target system can be expanded in terms of a complete set of the projectile states $\phi^{j}_{k}$. This involves a sum over the bound projectile states and an integral over momentum for the continuum states. In CDCC, the continuum is discretized and the total wave function is expanded as 
\begin{equation}
\Psi^{JM\pi} =\sum_{jkL} i^{L} \frac{u^{J\pi}_{jkL} \left(R\right)}{R} \left[ \phi^{j}_{k}\otimes Y_{L}\left( \Omega_{R} \right)\right]^{JM} ,
\label{totalpsi}
\end{equation} 
where L is the projectile-target angular momentum. For the systems under investigation, the convergence is reached for $L=200$. The summation (\ref{totalpsi}) is also truncated at a maximum angular momentum $j_{max}$, as well as at a maximum pseudostate energy, which limits k. 

The total Hamiltonian (\ref{totalH}) acting on Eq. (\ref{totalpsi}) leads to a set of coupled channel equations
\begin{align}
& -\frac{\hbar^{2}}{2\mu} \left[ \frac{d^{2}}{dR^{2}}  - \frac{L\left(L+1\right)}{R^{2}} \right] u^{J\pi}_{c}\left(R\right) + \nonumber\\ 
& \sum_{c^{\prime}}i^{L^{\prime}-L}V^{J\pi}_{cc^{\prime}}\left(R\right) u^{J\pi}_{c^{\prime}}\left(R\right) = \left(E-E^{j}_{k}\right) u^{J\pi}_{c}\left(R\right) ,
\label{Eqcc}
\end{align} 
where c indicates a $(jkL)$ channel. The relative energy $E$ is defined from the ground state of the projectile. $V^{J\pi}_{cc^{\prime}}$ potentials describe the nuclear and Coulomb couplings between the internal states $\phi^{j}_{k}$ of the projectile:
{\small\begin{align}
& V^{J\pi}_{cc^{\prime}}\left(R\right) = \nonumber\\
& \left\langle \left[ \phi^{j}_{k}\otimes Y_{L}\left( \Omega_{R} \right)\right]^{J}\left| \sum^{A}_{i=1} V_{ti}\left( \textbf{r}_{i}-\textbf{R} \right) \right| \left[ \phi^{j^{\prime}}_{k^{\prime}}\otimes Y_{L^{\prime}}\left( \Omega_{R} \right)\right]^{J} \right\rangle .
\label{Eq9}
\end{align}}
In the MCDCC calculation, the nucleon-nucleus optical potential $V_{ti}$ is independent of $L$, as the spin-orbit part of the potential is neglected.

By noting that 
\begin{align}
& \left\langle \phi^{j^{\prime}m^{\prime}}_{k^{\prime}}\left| \sum^{A}_{i=1} V_{ti}\left( \textbf{r}_{i}-\textbf{R} \right) \right|\phi^{jm}_{k} \right\rangle = \nonumber\\
& \left\langle \phi^{j^{\prime}m^{\prime}}_{k^{\prime}}\left| \int \sum^{A}_{i=1} V_{ti}\left(\textbf{r} \right) \delta\left( \textbf{r}_{i}-\textbf{R}-\textbf{r} \right) d^{3}\textbf{r} \right|\phi^{jm}_{k} \right\rangle  = \nonumber\\
& \int V_{ti}\left(\textbf{r}\right) \rho_{j^{\prime}m^{\prime}k^{\prime},jmk}\left(\textbf{r + R}\right) d^{3}\textbf{r},
\label{Eq12}
\end{align}
and by using a multipole decomposition, analogous to Eq. (\ref{Eqmulti}), for the proton and neutron densities, the radial matrix elements (\ref{Eq12}) are obtained in terms of $V^{(\lambda)}_{ti}$ multipoles, which are calculated by folding the projectile proton and neutron density multipoles  with the nucleon-target optical potential in the momentum space \cite{SaLo79}.  $\lambda_{max}=2$ is chosen for the multipole expansion in the calculations.

The coupled channel equations (\ref{Eqcc}) are solved using the $R$-matrix method on a Lagrange mesh \cite{He98,Dr10} (see the Appendix).  A channel radius $a=20$ fm is chosen for the systems with the $^{17}$F or $^{17}$O projectiles. The $R$ matrix is propagated over four intervals, each interval employing 150 Lagrange mesh points. For the $^{19}$F + $^{208}$Pb system, the chosen channel radius is 25 fm, and the $R$ matrix is propagated over four intervals, each employing 190 mesh points.
 
\subsection{Conditions of the calculation}
\subsubsection{Effective NN interaction}
The effective NN interaction employed in the MCDCC internal Hamiltonian (\ref{H0}) involves a central nuclear component, the Coulomb potential and a zero-range spin-orbit term as defined in Ref. \cite{DeDu11}. The noncentral forces, as well as the one pion exchange potential (OPEP) contributions, are simulated by an appropriate choice of central effective interaction, which is well adapted to harmonic oscillator orbitals.

Parameters of the effective NN interaction are adjusted to reproduce the excitation energies of the ground state and an excited state of the projectile. The NN interaction, based on a Volkov force, parametrized to reproduce the ground state and the first excited state of the $^{17}$F projectile, provides almost identical elastic scattering cross sections at 10 MeV/nucleon to those, that use the Minnesota NN interaction. As the sensitivity to the choice of the effective NN interaction is small, all calculations are performed with the Minnesota interaction.

The Minnesota interaction parameters, that reproduce the $^{17}$F bound states, are the admixture parameter u$=0.917$ and the spin-orbit amplitude S$_{0}=33.8$ MeV~fm$^{5}$. Parameters u$=0.9245$ and S$_{0}=34.3$ MeV~fm$^{5}$ reproduce the ground state and the first excited state of $^{17}$O. Negative-parity states of  $^{17}$O are not described by the $^{16}$O + n configuration, and, therefore, are among the pseudostates that simulate the continuum. The experimental transition probability $B(E2, 5/2^{+} \rightarrow 1/2^{+})$ from the first excited state to the ground state of $^{17}$F is 64.9 $e^{2}fm^{4}$, the corresponding value from the GCM calculation is 50.1 $e^{2}fm^{4}$. 

\begin{figure}[ht]
\includegraphics[width=8cm]{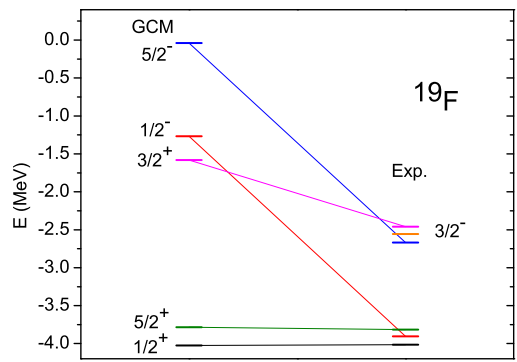}
\caption{\label{Fig11} (Color online)  Bound states of $^{19}$F with respect to the  $^{15}$N + $\alpha$ threshold, resulting from a Minnesota effective NN interaction with parameters u $=0.8185$ and S$_{0}=35.0$ MeV~fm$^{5}$.}
\end{figure}

For the $^{19}$F projectile, the parameters of the effective NN interaction, that reproduce the ground state (g.~s.), as well as the $5/2^{+}$ bound state, are u$=0.8185$ and S$_{0}=35.0$ MeV~fm$^{5}$. The negative-parity states as well the bound $3/2^{+}$ are much higher compared to the experimental energy levels, as shown in Fig. \ref{Fig11}. 
\subsubsection{Model space}
In the CDCC method, the total wave function (\ref{totalpsi}) is expanded in terms of the projectile states $\phi^{j}_{k}$. The projectile continuum is truncated at a maximum projectile excitation energy and a maximum angular momentum $j_{max}$. 

\begin{figure}[ht]
\includegraphics[width=8cm]{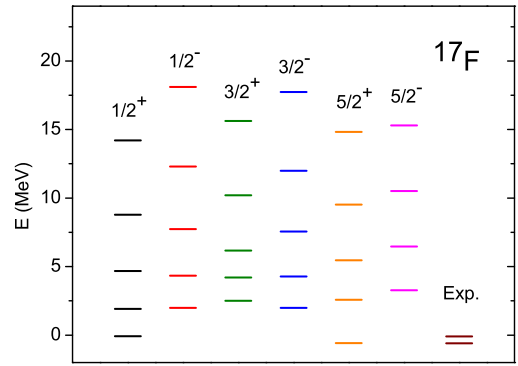}
\caption{\label{Fig9} (Color online)  $^{17}$F states included in the MCDCC calculation.}
\end{figure}

For the systems with $A=17$ projectile, the full calculation includes pseudostates up to 20 MeV. For the $^{17}$F projectile, the j$^{\pi}_{max}=5/2$ states, included in the calculation, are shown in Fig. \ref{Fig9}. For the $^{17}$O projectile, the angular momentum j$^{\pi}_{max}$ is cut at $7/2$ . Increasing the cut-off energy or $j_{max}$ of the projectile partial waves does not change the resulting cross sections.  

The calculation for the $^{19}$F + $^{208}$Pb system includes pseudostates up to 10 MeV. The cut-off angular momentum  j$^{\pi}_{max}$ is  $5/2$.

\subsubsection{Nucleon-target optical potential}
Ideally, one would want to find the energy-dependent nucleon-target interaction that reproduces the experimental cross sections without any further parameters. That would allow reliable predictions for the cross sections, where the experimental data are not available. 

\begin{figure}[ht]
\includegraphics[width=8cm]{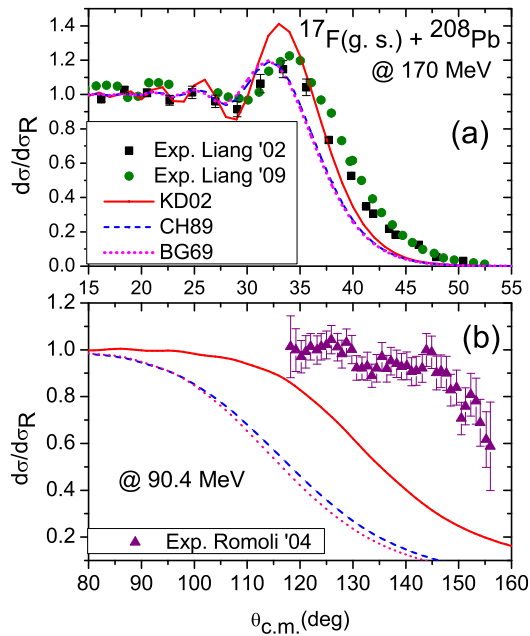}
\caption{\label{Fig3} (Color online)  Differential elastic cross sections, relative to Rutherford, for the scattering of $^{17}$F on $^{208}$Pb. The top panel and the bottom panel show the scattering at E$_{lab}=170$ MeV and at E$_{lab}=90.4$ MeV, respectively. The solid line shows the single-channel calculation that uses the Koning-Delaroche \cite{KD02} nucleon-target potential (KD02), the dashed line corresponds to the CH89 \cite{CH89} potential and the dotted line to the Bechetti-Greenlees (BG69)  \cite{BG69} potential. Quasielastic scattering data at E$_{lab}=170$ MeV and E$_{lab}=90.4$ MeV are from Ref. \cite{Li09} and Ref. \cite{Ro04}, respectively.}
\end{figure}

In Fig. \ref{Fig3}, we compare the single-channel $^{17}$F+$^{208}$Pb elastic scattering cross sections, resulting from three different global nucleon-target optical potentials \cite{KD02,CH89,BG69}. The spin-orbit part of all these potentials is not included in the calculations. The top panel of Fig. \ref{Fig3} shows the calculation at 10 MeV/nucleon. The single-channel calculations underestimate the cross sections at larger angles. An agreement to data is acceptable. At this energy, the Koning-Delaroche \cite{KD02}, the CH89 \cite{CH89} and the Bechetti-Greenlees \cite{BG69} interactions result in very similar elastic scattering cross sections. 

The bottom panel of Fig. \ref{Fig3} shows the $^{17}$F(g.~s.)+$^{208}$Pb elastic cross sections at E$_{lab}=90.4$ MeV. The single-channel calculations are in  poor agreement with the data. The difference between the Koning-Delaroche \cite{KD02} cross sections and the other two potentials \cite{CH89,BG69} is large. The CH89 \cite{CH89} potential only fits the data at or above 10 MeV/nucleon, to avoid the problems with compound elastic scattering \cite{HF52}. Ref. \cite{BG69} notes that substantial compound elastic scattering corrections are needed for the data below 10 MeV/nucleon. 

\begin{figure}[ht]
\includegraphics[width=8cm]{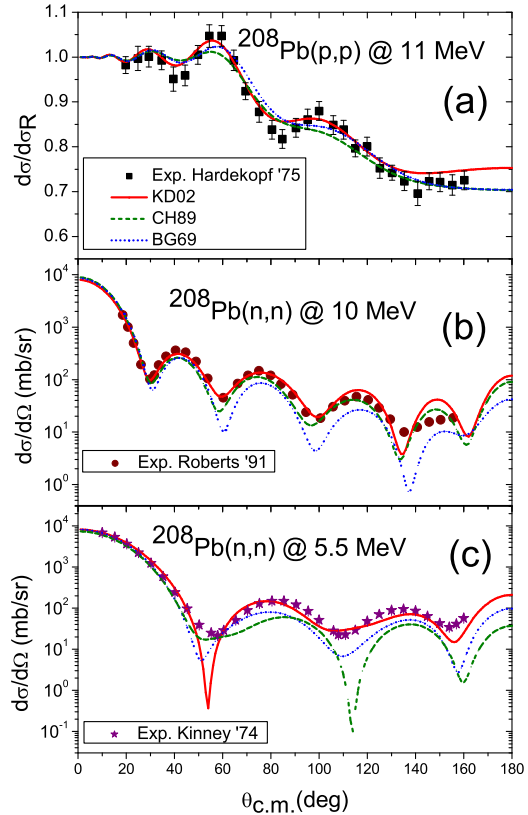}
\caption{\label{Fig4} (Color online)  The top panel shows the differential elastic cross sections, relative to Rutherford, for the proton scattering of $^{208}$Pb at 11 MeV. The middle panel and the bottom panel show the differential elastic cross sections for the neutron scattering of $^{208}$Pb  at 10 MeV and 5.5 MeV, respectively. The solid line corresponds to a Koning-Delaroche \cite{KD02} nucleon-target potential (KD02), the dashed line results from the CH89 \cite{CH89} potential and the dotted line is for the Bechetti-Greenlees (BG69)  \cite{BG69} potential.  The proton scattering data are taken from Ref. \cite{ET75}, and  the neutron scattering data are from Ref. \cite{Ro91} and \cite{ORNL} as in \cite{KD02} at 10 MeV and 5.5 MeV, respectively.}
\end{figure}

Figure \ref{Fig4} shows the proton and  neutron elastic scattering of $^{208}$Pb for the three aforementioned interactions. The top panel of Fig. \ref{Fig4} shows the proton scattering at 11 MeV, the middle panel shows the neutron scattering at 10 MeV, and the bottom panel shows the neutron scattering at 5.5 MeV. Results presented in the next section employ the Koning-Delaroche optical potential, which provides the best agreement with the data.
\section{Results}

\subsection{Application to $^{17}$F projectile}
In this section, the MCDCC method is applied to the $^{17}$F + $^{58}$Ni and $^{17}$F + $^{208}$Pb elastic scattering reactions. The Coulomb barrier is $V_{B,lab}\approx46$ MeV for the $^{58}$Ni target, as quoted in Ref. \cite{Ma10}, and $V_{B,lab}=91.7$ MeV for $^{208}$Pb target, as quoted in Ref. \cite{Si10}. 

\begin{figure}[ht]
\includegraphics[width=8cm]{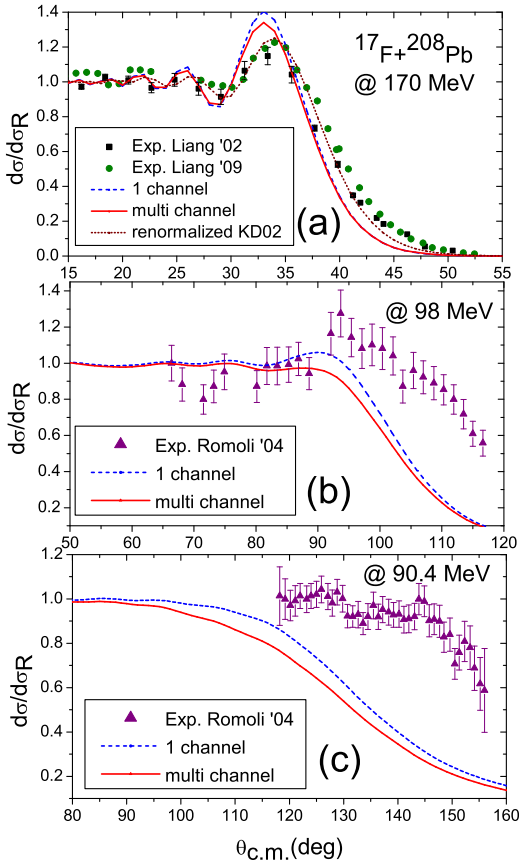}
\caption{\label{Fig17}  (Color online)  Differential elastic cross sections, relative to Rutherford, for the scattering of $^{17}$F on $^{208}$Pb. The top panel shows the scattering at 10 MeV/nucleon. The solid line shows the full MCDCC calculation and the dashed line shows the single-channel $^{17}$F+$^{208}$Pb calculation. The dotted line shows a full MCDCC calculation with a renormalized Koning-Delaroche potential. The scaling factor for the real part of the potential is 0.65. Quasielastic scattering data are taken from Ref. \cite{Li09}. The middle panel and the bottom panel show the cross sections for the same system at 98 MeV and 90.4 MeV, respectively. The solid line shows the full MCDCC calculation and the dashed line shows a single-channel $^{17}$F+$^{208}$Pb calculation. Quasielastic scattering data are taken from Ref. \cite{Ro04}.}
\end{figure}

For the $^{17}$F + $^{208}$Pb system, well above the Coulomb barrier, at E$_{lab}=170$ MeV, the calculated differential elastic scattering cross sections, relative to Rutherford cross section, are shown in the top panel of  Fig. \ref{Fig17}. The figure shows the single-channel as well as the multichannel calculations. The single-channel and the two-channel $^{17}$F(g. s., first excited state) cross sections are indistinguishable at the scale of the figure. The role of excited channels is small. This result is consistent with the conclusion of Ref. \cite{KuMo12}, in which the difference between the single-channel and a full calculation is not significant \cite{Mo}.

\begin{figure}[ht]
\includegraphics[width=8cm]{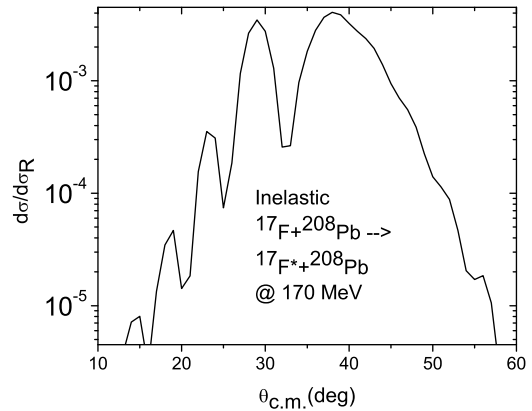}
\caption{\label{Fig10} The inelastic cross section, relative to Rutherford, for the scattering of $^{17}$F on $^{208}$Pb at 10 MeV/nucleon.}
\end{figure}

The cross sections at 170 MeV are compared to the quasielastic data from Ref. \cite{Li09} as in Ref. \cite{KuMo12}, in which the $1/2^{+}$ excited state could not be separated from the data. As shown in Fig. \ref{Fig10}, the theoretical contribution from the inelastic cross sections due to the first excited state of $^{17}$F is small, leading to quasielastic cross sections almost identical to the elastic scattering cross sections. 

\begin{figure}[ht]
\includegraphics[width=8cm]{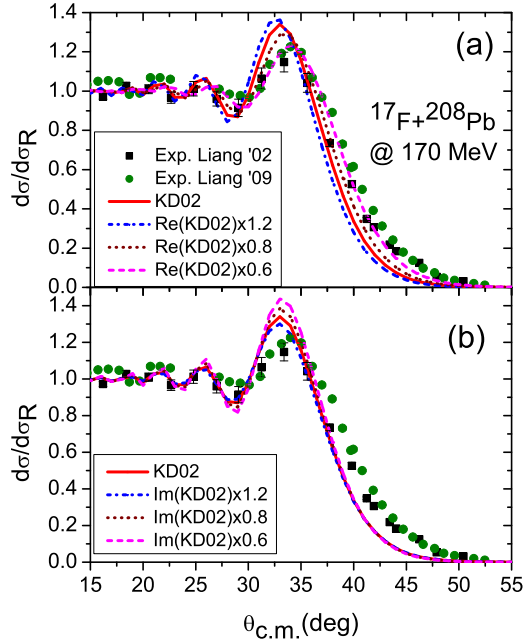}
\caption{\label{Fig2} (Color online)  Differential elastic cross sections, relative to Rutherford, for the scattering of $^{17}$F on $^{208}$Pb at 10 MeV/nucleon. The top (bottom) panel shows the full MCDCC calculation with the real (imaginary) part of the Koning-Delaroche \cite{KD02} potential scaled by 1.2, 1.0, 0.8 and 0.6. The quasielastic scattering data are taken from Ref. \cite{Li09}.}
\end{figure}

To better reproduce the data, the real part of the optical Koning-Delaroche potential needs to be renormalized by a 0.65 factor. The effect of a renormalization on the $^{17}$F+$^{208}$Pb elastic cross section at 10 MeV/nucleon is illustrated in Fig. \ref{Fig2}, by increasing and reducing only the real part (top panel) and only imaginary part (bottom panel) of the Koning-Delaroche potential.

Just above and below the Coulomb barrier, at E$_{lab}=98$ MeV and E$_{lab}=90.4$ MeV, the $^{17}$F + $^{208}$Pb elastic scattering cross sections, relative to Rutherford, are shown in the middle and the bottom panels of Fig. \ref{Fig17}, respectively. The agreement with the data is poor. Similar results are obtained in the traditional  CDCC \cite{Mo}. Experimental data at lower energies are taken from Ref. \cite{Ro04}. The contributions from the first excited states in both $^{17}$F and $^{208}$Pb are not separated from the experimental scattering data. In the MCDCC calculation, the $^{208}$Pb target is structureless and the inelastic scattering contribution from the $1/2^{+}$ excited state in $^{17}$F is small. Therefore, the elastic scattering cross sections are indistinguishable from the quasielastic scattering cross sections.

\begin{figure}[ht]
\includegraphics[width=8cm]{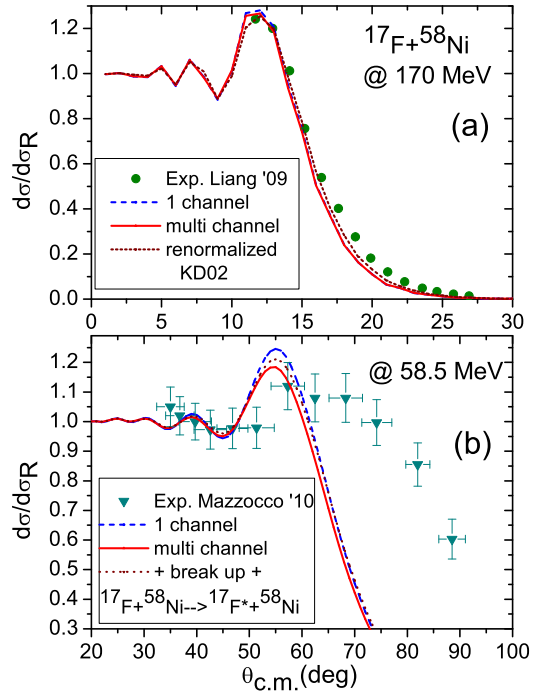}
\caption{\label{Fig18} (Color online)  Differential elastic cross sections, relative to Rutherford, for the scattering of $^{17}$F on $^{58}$Ni. The top panel shows the scattering at 10 MeV/nucleon. The solid line shows the full MCDCC calculation and the dashed line shows the single-channel $^{17}$F+$^{208}$Pb calculation. The dotted line shows a full MCDCC calculation with a renormalized Koning-Delaroche potential. The scaling factor for the real part of the potential is 0.8. Quasielastic scattering data are taken from Ref. \cite{Li09}. The  bottom panel shows the scattering at 58.5 MeV. The solid line shows the full MCDCC calculation, the dashed line shows the single-channel $^{17}$F+$^{208}$Pb calculation and the dotted line shows the cross sections that include the inelastic and breakup contributions. Quasielastic scattering data are taken from Ref. \cite{Ma10}.}
\end{figure}

For the $^{17}$F + $^{58}$Ni system, well above the Coulomb barrier, at E$_{lab}=170$ MeV, the calculated differential elastic cross sections, relative to Rutherford cross section, are shown in the top panel of  Fig. \ref{Fig18}. The couplings to continuum are small. The elastic cross sections are compared with the quasielastic data from Ref. \cite{Li09} as in Ref. \cite{KuMo12}, in which the $1/2^{+}$ excited state could not be separated from the data.  Without any additional parameter, the MCDCC provides fair cross sections. The calculated contribution from the inelastic cross sections due to the first excited state of $^{17}$F is small, leading to quasielastic cross sections indistinguishable from the elastic scattering cross sections. To better reproduce the experimental cross sections, the real part of the optical Koning-Delaroche potential can be renormalized by a factor of 0.8. 

The quasielastic experimental data for the $^{17}$F + $^{58}$Ni reaction just above the Coulomb barrier at 58.5 MeV \cite{Ma10} include the inelastic processes leading to an excitation of the projectile and the target low-lying states, transfer reaction products and $^{16}$O breakup fragments. As those contributions compared to elastic scattering cross sections are small \cite{Ma10}, a comparison to the MCDCC elastic scattering cross sections is justified. The single-channel cross sections, the full MCDCC calculation, and the cross sections that include the inelastic as well as breakup contributions, are shown in the lower panel of Fig. \ref{Fig18}. 

\subsection{Application to $^{17}$O projectile}
\begin{figure}[ht]
\includegraphics[width=8cm]{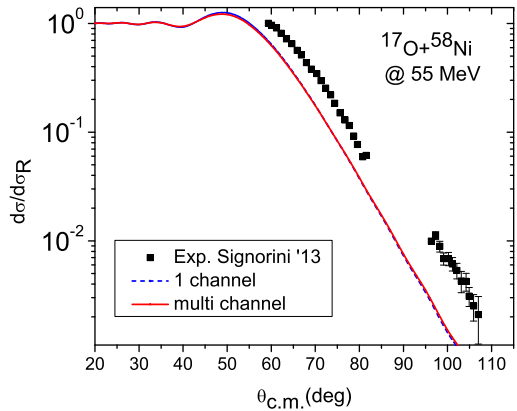}
\caption{\label{Fig8} (Color online)  Differential elastic cross sections, relative to Rutherford, for the scattering of $^{17}$O on $^{58}$Ni at E$_{lab}=55$ MeV. The solid line shows the full MCDCC calculation and the dashed line shows the single-channel $^{17}$O+$^{58}$Ni calculation. Quasielastic experimental data are taken from Ref. \cite{Si13}.}
\end{figure}

The Coulomb barrier for the $^{17}$O + $^{208}$Pb system is $V_{B,lab}\approx83$ MeV, as quoted in Ref. \cite{Ro04}, and $V_{B,lab}\approx44$ MeV for the $^{17}$O + $^{58}$Ni system. The differential elastic scattering cross sections of the well bound $^{17}$O nucleus (S$_{n}=4.1$ MeV) on $^{58}$Ni just above the Coulomb barrier at E$_{lab}=55$ MeV, relative to Rutherford cross section, are shown in Fig. \ref{Fig8}. The single-channel and the multichannel calculations provide very similar cross sections on a logarithmic scale. The role of excited states is again small. The MCDCC calculation underestimates the cross sections at larger angles. Quasielastic experimental data \cite{Si13} do not exclude the contribution from the $1/2^{+}$ excited state of $^{17}$O. As in the case of the $^{17}$F projectile, the theoretical inelastic scattering cross section to the first excited state of $^{17}$O is small.

\begin{figure}[ht]
\includegraphics[width=8cm]{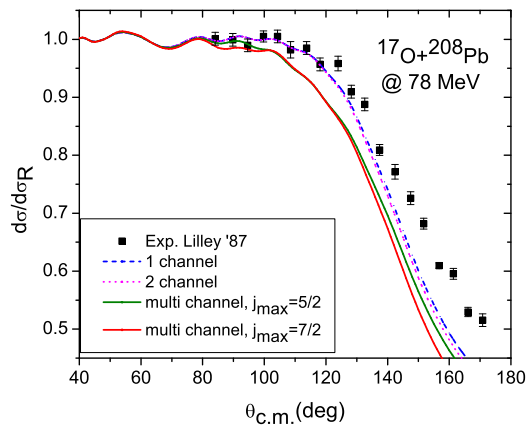}
\caption{\label{Fig7} (Color online)  Differential elastic cross sections, relative to Rutherford, for the scattering of $^{17}$O on $^{208}$Pb at E$_{lab}=78$ MeV. The solid line shows the full MCDCC calculation, the dashed line and the dotted line show a single-channel $^{17}$O+$^{208}$Pb and a two-channel $^{17}$O(g.~s., 1st ex.~s.) + $^{208}$Pb calculation, respectively. Experimental data are taken from Ref. \cite{Li87}.}
\end{figure}

Just below the Coulomb barrier, at E$_{lab}=78$ MeV, for the $^{17}$O + $^{208}$Pb system, the differential elastic cross sections, relative to Rutherford cross section, are shown in Fig. \ref{Fig7}. In this case, the excited states of $^{17}$O are contributing significantly. Figure \ref{Fig7} shows the single-channel, the two-channel and the multichannel calculations  without a renormalization factor for the Koning-Delaroche potential. Experimental data are taken from Ref. \cite{Li87}. Unlike in the case of $^{17}$F, for which the agreement with data is poor near the Coulomb barrier, the elastic scattering cross sections obtained for the $^{17}$O projectiles are closer to the data. 

\subsection{Application to $^{19}$F projectile}

The Coulomb barrier for the $^{19}$F + $^{208}$Pb system is  $V_{B,lab}\approx94$ MeV. The calculated cross sections are weakly sensitive to E$_{max}$ and hence to the quality of the spectrum.

\begin{figure}[ht]
\includegraphics[width=8cm]{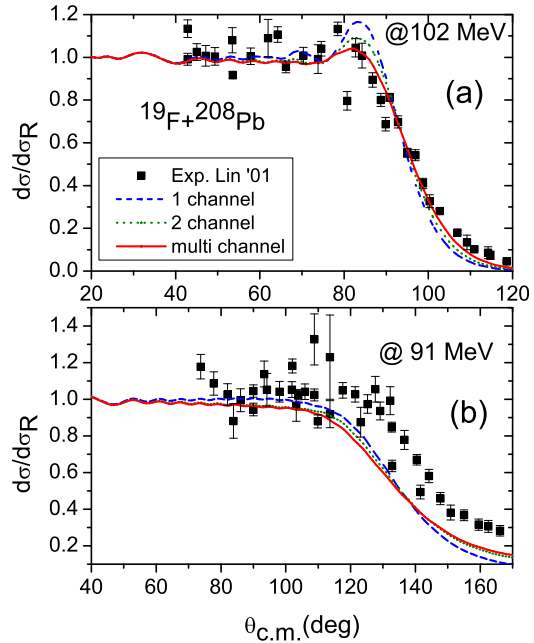}
\caption{\label{Fig16} (Color online)  Differential elastic cross sections, relative to Rutherford, for the scattering of $^{19}$F on $^{208}$Pb. The top panel and the bottom panel show scattering at E$_{lab}=102$ MeV and E$_{lab}=91$ MeV, respectively. The dashed line shows the single-channel $^{19}$F+$^{208}$Pb calculation, the dotted line shows the two-channel $^{19}$F(g.~s., $5/2^{+}$) + $^{208}$Pb calculation and the solid line shows the full MCDCC calculation. Quasielastic scattering experimental data are taken from Ref. \cite{Li01}.}
\end{figure}

The differential elastic scattering cross sections, relative to Rutherford, just above the Coulomb barrier at 102 MeV and just below the Coulomb barrier at 91 MeV are shown in the top panel and the bottom panel of Fig. \ref{Fig16}, respectively.
Figure  \ref{Fig16} shows the single-channel $^{19}$F(g.~s.) + $^{208}$Pb, the two-channel $^{19}$F(g.~s., $5/2^{+}$) + $^{208}$Pb and the full MCDCC calculations. The cross sections at 102 MeV, shown in the top panel of Fig. \ref{Fig16},  are in reasonable agreement with the data without any renormalization factor. 

At 91 MeV, the calculations, shown in the bottom panel of Fig.  \ref{Fig16}, underestimate the cross sections at large angles. The calculations do not include any renormalization factor for the nucleon-target optical potential.  Quasielastic scattering data \cite{Li01} do not exclude the inelastic scattering due to the $1/2^{-}$ and $5/2^{+}$ excited states. The theoretical inelastic contribution to the quasielastic scattering data is small. 

\section{Conclusions}
The MCDCC is very satisfactory in the sense, that it includes the microscopic properties of stable and exotic nuclei in a reaction model. The structure of the fully antisymmetric projectile is described via an effective NN interaction. The nucleon-target interaction in a folding model relies on the well known  nucleon-nucleus optical potentials. The target in the model is structureless, and the cross sections are weakly sensitive to the choice of the Volkov or the Minnesota NN force.

The MCDCC could be improved in several ways. It can include the target, as well as the additional core excitations. At lower energies, the role of the target excitations is expected to be small. The present calculations treat $^{17}$F as a $^{16}$O + $p$ configuration, $^{17}$O as $^{16}$O + $n$, and $^{19}$F as  $^{14}$N + $\alpha$. Other configurations can also be coupled, e.g. $^{13}$N + $\alpha$ for $^{17}$F, however that would be more complicated.

At 170 MeV, the cross sections for the $^{17}$F + target systems are in a good agreement with data. Just above and below the Coulomb barrier, an agreement with the data for the $A=17$ projectiles is much less satisfactory. The reasons for this discrepancy, which is also found in the traditional CDCC, are unclear. For all systems under investigation, the breakup and the inelastic scattering are small and thus cannot account for the difference in the cross sections at near-Coulomb energies. A good agreement with the data  is reached for the $^{19}$F + $^{208}$Pb system at 102 MeV, which is just above the Coulomb barrier; however, the cross sections are underestimated just below the Coulomb barrier, at 91 MeV. 

In most assessed cases, the coupling to continuum is not significant. A substantial contribution from the excited channels is found for the elastic scattering of $^{17}$O + $^{208}$Pb system at 78 MeV.
\section*{Acknowledgments}
The authors thank A.~M. Moro for the discussions and the unpublished results on traditional CDCC calculations. This text presents research results of the IAP program P7/12, initiated by the Belgian-state Federal Services for Scientific, Technical and Cultural Affairs. P.~D. acknowledges the support of the National Fund for Scientific Research (FNRS), Belgium.

\appendix*
\section{$R$-matrix method on a Lagrange mesh}
The $R$-matrix formalism divides the configuration space into internal and external regions, separated by the boundary surface, defined by channel radii. These radii are chosen large enough, so that the system would interact only through the known long range forces in the exterior region. The scattering wave functions in the asymptotic region are approximated by their asymptotic expressions, which are known, except for the coefficients related to the scattering matrix U$_{cc^{\prime}}$. In the interior region, the radial wave functions $u^{J\pi}_{c}$ in Eq. (\ref{Eqcc}) are expanded over a discrete basis
\begin{equation}
u^{J\pi}_{c}\left(R\right) =\sum^{N}_{n=1} A^{J\pi}_{cn} \varphi_{n}\left(R\right).
\label{Eq13}
\end{equation}

In the Lagrange mesh method, ${\varphi_{i}}\left(x\right)$ are the Lagrange basis functions, associated with the Gauss quadrature \cite{He98}. Every function ${\varphi_{i}}\left(x\right)$ satisfies Lagrange conditions ${\varphi_{j}}\left(x_{i}\right)=\lambda^{-1/2}_{i} \delta_{ij}$, that is it vanishes at all mesh points except one. $x_{i}$ are the $N$ roots of the associated orthogonal polynomial and $\lambda_{i}$ are the weights of the Gauss quadrature formula \cite{He98,Dr10}. The here employed method  relies on a mesh of points $\left\{x_{i}\right\}$ in the interval $[0, a]$ built with the zeros of a Legendre polynomial of degree $N$,  $P_{N}\left(2x_{n}-1\right)=0$  \cite{Dr10}. 

The scattering matrix U$_{cc^{\prime}}$  and the expansion coefficients $A^{J\pi}_{cn}$ in Eq. (\ref{Eq13}) are determined by matching solutions in the internal and external regions. By using a common channel radius $a$ for all channels, the $R$ matrix is expressed as in Ref. \cite{Dr10} as
\begin{equation}
R^{J\pi}_{cc^{\prime}}= \frac{\hbar^{2}}{2\mu a} \sum_{nn^{\prime}} \varphi_{n} \left(a\right) \left[ \left(\textbf{C}^{J\pi}\right)^{-1} \right]_{cn,c^{\prime}n^{\prime}}  \varphi_{n^{\prime}} \left(a\right)  ,
\end{equation}
{\small\begin{equation}
C^{J\pi}_{cn,c^{\prime}n^{\prime}}=\left\langle \varphi_{n} \left| \left(T+\mathcal{L}+E^{j}_{k}-E\right) \delta_{cc^{\prime}}+V^{J\pi}_{cc^{\prime}} \right| \varphi_{n^{\prime}} \right\rangle  .
\label{Eqc}
\end{equation}}
Here, the Bloch operator $\mathcal{L}$ in Eq. (\ref{Eqc}) is added to make the Hamiltonian (\ref{totalH}) in Eq. (\ref{Eqcc}) Hermitian. In the Lagrange mesh method, the calculation of $C^{J\pi}_{cn,c^{\prime}n^{\prime}}$ matrix elements (\ref{Eqc}) is limited to the mesh points. By using the Gauss approximation, the overlap between the Lagrange basis functions is $\left\langle \varphi_{i} | \varphi_{j} \right\rangle =  \delta_{ij}$, the matrix elements for the potential are $\left\langle \varphi_{i}\left| V \right|\varphi_{j} \right\rangle = V\left(ax_{i}\right)\delta_{ij}$, and the analytic expressions for the matrix elements of the kinetic energy and the Bloch operator at mesh points are given in Ref. \cite{He98}.

The collision matrix U$_{cc^{\prime}}$ is defined as  \cite{Dr10}
\begin{equation}
\textbf{U}^{J\pi}=\left(\textbf{Z}^{J\pi}_{O}\right)^{-1} \textbf{Z}^{J\pi}_{I},
\end{equation}
where 
{\small\begin{equation}
\left(Z^{J\pi}_{O}\right)_{cc^{\prime}} = \left(k_{c^{\prime}}a\right)^{-1/2}\left[O_{L}\left(k_{c}a\right)\delta_{cc^{\prime}}- k_{c^{\prime}}a R^{J\pi}_{cc^{\prime}}O^{\prime}_{L^{\prime}}\left(k_{c^{\prime}}a\right)\right] .
\label{Eq25}
\end{equation}}
$\left(Z^{J\pi}_{I}\right)_{cc^{\prime}}$ is defined by replacing the outgoing waves $O_{L}$ and derivatives in Eq. (\ref{Eq25}) by incident waves $I_{L}$ and their derivatives.

To improve the efficiency of the $R$-matrix calculation, the model employs the propagation method, briefly presented in Ref. \cite{DeBa10}. The wave function of the $R$ matrix is propagated over several intervals, thereby reducing the required size of the basis $\left\{\varphi_{i}\right\}$.

\end{document}